\documentclass[12pt]{article}
\usepackage{amsmath,amsthm,amsfonts,amssymb,amscd}
\usepackage{graphics}
\usepackage[latin1]{inputenc}

\renewcommand{\thefootnote}

\begin{document}

\centerline{\large\bf A note on Feynman Path Integral}

\smallskip

\centerline{\large\bf for Electromagnetic External Fields}

\vglue .4in

\centerline{\large\bf Luiz C.L. Botelho}

\vglue .3in

\centerline{Departamento de Matemática Aplicada}

\smallskip

\centerline{Instituto de Matemática, Universidade Federal Fluminense}

\smallskip

\centerline{Rua Mario Santos Braga, CEP 24220-140}

\smallskip

\centerline{Niterói, Rio de Janeiro, Brasil}

\smallskip

\centerline{e-mail: botelho.luiz@superig.com.br}

\vglue .5in

\noindent{\large\bf Abstract:} We propose a Fresnel stochastic white noise framework to analyze the nature of the Feynman paths entering on the Feynman Path Integral expression for the Feynman Propagator of a particle quantum mechanically moving under an external electromagnetic time-independent potential.

\medskip

\noindent{\large\bf Key words:} Feynman Path Integral, Nelson  Stochastic Mechanics, Stochastic Calculus.

\vglue .2in

\noindent{\large\bf 2.\, The Fresnel stochastic nature of Feynman path integral}

\bigskip

Let us start our note by considering as a basic object associated to a quantum particle of mass $m$, the following white noise functional Fresnel-Feynman path integral defined on an ensemble of closed, quantum trajectories of a white noise process with correlation function depending solely on the particle classical mass particle
\begin{equation}
I_m[j] = \int_{\vec{n}(\sigma)=0}^{\vec{n}(T)=0} D^F[\vec{n}(\sigma)] \exp\left(im \int_0^T \frac 12\,(\vec{n}(\sigma)^2 d\sigma\right) 
\exp\left(i \int_0^T \vec j(\sigma)\vec n(\sigma)d\sigma\right) \tag{1}
\end{equation}

Where $\vec n(\sigma)$ are the quantum white-noise Feynman closed trajectories on $R^3$ for the propagation time interval $\sigma \in [0,T]$ and satisfying the Dirichlet condition $\vec n(0) = \vec n(T) = 0$. Here $j(\sigma)$ denotes a fixed external real valued path source.

It is worth to call the reader attention that the normalized white-noise external source Feynman path integral eq(1) can be straightforwordly (heuristically) evaluated through a random Fourier series expansion for the random white noise trajectory $\vec n(\sigma)$\, $0 \le \sigma < T$ yelding the exact result:
\begin{equation}
\frac{I_m[\vec j]}{I_m[0]} = \exp\left\{\frac{i}{2m} \int_0^T (\vec j(\sigma))^2\,d\sigma\right\} \tag{2}
\end{equation}

In order to analyze the stochastic nature of the Feynman integral associated to the quantum system defined by a particle (of a fixed newtonian mass $m$)${}^{(1)}$ under the presence of an external time-independent electromagnetic field we firstly consider the well-defined unique system classical trajectory connecting the  spatial points $\vec x_1$ and $\vec x_2$ in time $T$.
\footnote{${}^{(1)}$The same mass parameter of the (vectorial) white noise eq(1).}

\noindent Namely:
\begin{equation}
\begin{aligned}
\frac{m\,d^2\vec{x}(\sigma)}{d\sigma^2} &= e\left(-\vec\nabla.\phi + \frac{d\vec x}{d\sigma} \times \vec \nabla \times \vec A\right)(\vec{x}(\sigma))\\
\vec{x}(0) &= \vec x_1\\
\vec{x}(T) &= \vec x_2 
\end{aligned} \tag{3}
\end{equation}

We now introduce what we call the effective classical potential, throught the Taylor expansion below defined
\begin{equation}
\hbar \,\phi_{eff}\big(x,[x^{cL}(\sigma)]\big) \overset{\rm definition}{\equiv} e\Phi\big(x^{cL}(\sigma)+\sqrt{\hbar}\,x\big) - e\Phi(x^{cL}(\sigma)) - \sqrt{\hbar}\big([\vec\nabla_\ell(e\phi)(x^{cL}(\sigma)]\times\vec x^\ell  \tag{4}
\end{equation}

Note that this effective classical potential also depends functionally on the system's classical trajectory $\{x^{cL}(\sigma)\}_{0\le\sigma\le T}$ and the Plank constant.

We now introduce the Feynman quantum trajectories of our system defined mathematically as those paths $\vec{x}^q(\sigma)$, functionals of the Feynman white-noise path $\vec{n}(\sigma)$, through the Hamilton-Jacobi equation for the quantum trajectory ([1]), where $E$ denotes the classical system total energy${}^{(2)}$

\footnote{${}^{(2)}$\,$\hbar\big(\vec{A}_i^{eff}(\vec x, [\vec{x}^{cL}(\sigma)])\big) = \vec{A}_i\big(\vec{x}^{cL}(\sigma) + \sqrt{\hbar} \vec x\big)-\vec{A}_i(\vec{x}^{cL}(\sigma) - \sqrt{\hbar}\left[\sum\limits_{k=1}^3 \left(\dfrac{\partial}{\partial x^k}\,\vec{A}_i\right)(\vec{x}^{cL}(\sigma))x^k\right]$
}
\begin{equation}
\frac{1}{2m} \left[\left(\vec\nabla.W^{eff} - \frac ec\,\vec {A}^{eff}\right)(\vec x,\vec{x}^{cL}(\sigma)\right]^2 + \bar{V}_{eff}(x,[\vec{x}^{cL}(\sigma]) = E \tag{5}
\end{equation}
Namelly:
\begin{equation}
\begin{aligned}
\frac{d\vec{x}^q(\sigma)}{d\sigma} &= \left(\frac 1m \left(\vec\nabla W^{eff} - \frac ec\, 
\vec{A}^{eff}\right)(\vec{x}^q(\sigma), [x^{cL}]) + \vec n(\sigma)\right)\\
\vec{x}^q(0) &= \vec{x}^q(T) = 0\\
\vec n(0) &= \vec n(T) = 0
\end{aligned} \tag{6}
\end{equation}

We claim that the full Feynman path defined below as quantum fluctuations around the classical path with ``size" of order $(\hbar)^{1/2}$,
\begin{equation}
\vec x(\sigma)=\vec{x}^{cL}(\sigma) + \sqrt{\hbar}(\vec{x}^q(\sigma)), \tag{7}
\end{equation}
can be mathematical used to be the set of paths that enters in the Feynman path integral expression for the quantum mechanical propagator, and leading straighforwardly to the expected result that on the asymptotic semi-classical limit $\hbar \to 0$, the leading contribution comes solely from the classical path.

To show these results, we firstly consider the formal object written fully below:
\begin{align*}
G(x_1,x_2,T) &= \exp\left(\frac{1}{\hbar}\, S[x^{cL}(\sigma)]\right)\times \exp\left\{-\frac{i}{\hbar} \int_0^T \frac{dx^{i,cL}(\sigma)}{d\sigma}\,\mathcal{E}_{ijk}\,
\frac{\partial}{\partial x^j}\,A^k(x^{cL}(\sigma))d\sigma\right\}\\
&\times \bigg\{\int_{x^q(0)=0}^{x^q(0)=0} D^F[\vec{x}^q(\sigma)]\bigg[\int_{\vec{n}(0)=0}^{\vec n(T)=0} D^F(\vec n(\sigma) \exp\left(im \int_0^T \frac 12\,(\vec n(\sigma))^2\,d\sigma\right)\bigg]\\
&\times {\rm det}_F\left[\frac{d}{d\bar\sigma} - \frac 1m\,\vec\nabla_x\left(\frac{\delta}{\delta\vec{x}_q(\bar\sigma)}\, W^{eff}\right) + \frac{\delta}{\delta\vec{x}_q(\sigma)}\left(\frac{ec}{m}\,\vec {A}^{eff}\right)\right]\\
&\times\delta^{(F)} \left[\frac{d\vec{x}^q(\sigma)}{d\sigma} - \frac 1m \left(\vec\nabla\,W^{eff} - \frac ec\,\vec {A}^{eff}\right)(\vec{x}^q(\sigma), [\vec{x}^{cL}(\sigma)]) - \vec n(\sigma)\right]\bigg\}\\
& \times (\exp(-iET)) \tag{8}
\end{align*}

We will heuristically show that it satisfies the system's quantum Feynman propagator equation:
\begin{equation}
 + i\hbar\,\frac{\partial}{\partial T}\,G(\vec{x}_1,\vec{x}_2,T) = \left(\frac{(-i\hbar \vec\nabla-\frac ec\,\vec A)^2}{2m} + e\,\phi\right)G(\vec{x}_1,\vec{x}_2,T) \tag{9-a}
\end{equation}
\begin{equation}
G(\vec{x}_1,\vec{x}_2,T) = \delta^{(3)}\,(\vec{x}_1-\vec{x}_2) \tag{9-b}
\end{equation}
\begin{equation}
({\rm div}\,\vec A)(\vec x) \equiv 0 \tag{9-c}
\end{equation}

In order to arrive at such result, we firstly compute exactly the white-noise path integral with the following result (rewritten entirely on term of the quantum path $\vec{x}^q(\sigma)$)
\begin{align*}
&G(\vec{x}_1,\vec{x}_2,T)= \exp(iET) \times \exp\left(\frac{i}{\hbar}\,S(x^{cL}(\sigma)\right)\\
&\times \left\{\int_{\vec{x}^q(0)=0}^{\vec{x}^q(T)} D^F[\vec{x}^q(\sigma)] \times \exp\left(\frac{im}{2} \int_0^T \left[\frac{d\vec{x}^q}{d\sigma} - \frac 1m \left(\vec\nabla W^{eff}-\frac ec\,\vec {A}^{eff}\right)^2(\vec{x}^q,\vec{x}^{cL})\right]^2d\sigma\right)\right\}\\
&\times {\rm det}_F \left[\frac{d}{d\bar\sigma} - \frac 1m\,\nabla_x \left(\frac{\delta}{\delta x_q(\bar \sigma)}\,W^{eff}\right) + \frac{\delta}{\delta\vec{x}_q(\sigma)} \left(\frac{ec}{m}\,\vec {A}^{eff}\right)\right] \tag{10}
\end{align*}
The action functional on the Feynman path integral thus posseses the following explicit functional form
\begin{align*}
im \int_0^T\frac 12 \bigg[\left(\frac{d\vec{x}^q}{d\sigma}\right)^2 &- \frac 2m\, \frac{d\vec{x}^q}{d\sigma}\,\left(\vec\nabla\,W^{eff} - \frac ec\,\vec{A}^{eff}\right)\\
&+ \frac{1}{m^2} \left(\left(\vec\nabla\,W^{eff} - \frac ec\,\vec{A}^{eff}\right)\right)^2\bigg](\sigma)d\sigma
\end{align*}
\begin{align*}
=\left(\frac{im}{2} \int_0^T \left(\frac{d\vec{x}^q}{d\sigma}\right)^2 (\sigma)d\sigma\right) + \bigg(\frac{im}{2} \int_0^T\bigg(-\frac 2m\, \frac{d\vec x}{d\sigma}\cdot\vec\nabla\,W^{eff}
- \frac{2e}{mc}\,\frac{d\vec{x}^q}{d\sigma}\cdot\vec {A}^{eff}\bigg)(\sigma)d\sigma
\end{align*}
\begin{align*}
+\bigg(\frac{im}{2} \int_0^T \frac{1}{m^2} \bigg((\vec\nabla\,W^{eff})^2 &+ \frac{e^2}{c^2}\,(\vec{A}^{eff})^2 - \frac{2e}{c}\,\vec {A}^{eff}\cdot\vec\nabla W^{eff}\bigg)(\sigma)d\sigma\bigg)\\
\tag{11}
\end{align*} 

By noting the result
\begin{equation}
\left(\int_0^T \left(\frac{d\vec{x}^q}{d\sigma}\cdot \vec\nabla\,W^{eff}\right)(\sigma)d\sigma\right) = 0 \tag{12}
\end{equation}
togheter with the delta function distribution prescription at coincident points below to make the functional determinant equals to $1$
\begin{equation}
\mathcal O(0)=0. \tag{13}
\end{equation}
We have thus the result
\begin{align*}
G(x_1,x_2,T) &= \exp\left(\frac{i}{\hbar}\, S[\vec{x}^{cL}(\sigma)]\right)\\
&\times \bigg\{\int_{\vec{x}^q(0)=0}^{\vec{x}^q(T)=0} D^F[\vec{x}^q(\sigma)] \left(\exp i \int_0^T \left(\frac{m}{2}\,\frac{dx^q}{d\sigma}\right)^2 (\sigma) d\sigma\right)\\
&\times \exp\left(-\frac{ie}{c}\left(\int_0^T \vec{A}^{eff}(\vec{x}^q,[\vec{x}^{cL}(\sigma)]) \frac{d\vec{x}^q}{d\sigma}\,(\sigma)d\sigma\right)\right)\\
&\times \exp\left(-ie \int_0^T \phi_{eff} (x^q(\sigma))d\sigma\right)\bigg\}\tag{14}
\end{align*}

We now observe that through the classical motion equation eq(3) and the definition of the effective potential eq(4) and the formal invariance of translation under classical trajectories of the Feynman path measure (see ref[2]) one can re-write eq(14) in the usual Feynman form
\begin{equation}
G(\vec{x}_1,\vec{x}_2,T) = \int_{\vec x(0)=\vec{x}_1}^{\vec x(T)=\vec{x}_2} D^F[\vec x(\sigma)]\, \exp\left\{\frac{i}{\hbar}\, S[\vec x(\sigma)]\right\}, \tag{15}
\end{equation}
with the classical system action${(3)}$

\footnote{${}^{(3)}$To arrive at the final result eq(16) we suppose the ``orthogonality" conditions among the classical and quantum (fluctuation) Feynman path when in presence of an external time-independent classical external electromagnetic field: 
\begin{align*}
&\frac{ie}{c\sqrt{\hbar}}\left(\int_0^T \frac{d\vec{x}_i^{cL}}{d\sigma}\cdot\frac{\partial A^i}{\partial x^\ell}\, (\vec{x}^{cL}) \vec{x}^{q,\ell}(\sigma)d\sigma\right)=0\\
&\frac{ie}{c} \left(\int_0^T \frac{d\vec{x}_i^{cL}(\sigma)}{d\sigma}\, A^{i,eff}(\vec{x}^q, \vec{x}^{cL}(\sigma))d\sigma\right)=0\\
&\frac{ie}{c\sqrt\hbar} \left(\int_0^T \frac{d\vec{x}_i^q}{d\sigma}\, A^i(\vec{x}^{cL}(\sigma)d\sigma\right)=0\\
&\frac{ie}{c} \left(\int_0^T \frac{d\vec{x}^{q,i}}{d\sigma}\cdot\frac{\partial A_i}{\partial x^\ell}\,(\vec{x}^{cL}(\sigma)).\vec{x}^{q,\ell}(\sigma)d\sigma\right)=0
\end{align*}

Otherwise we have all of ours results above written holding true for the quantum closed trajectories ($\vec{x}^{cL}(\sigma)\equiv 0$), which is surely relevant at the level of the quantum statistical canonical partition functional or to the trace of the Feynman propagator.}

\begin{align*}
S[\vec x(\sigma)] = \frac{m}{2} \int_0^T \left(\frac{d\vec x}{d\sigma}\right)^2(\sigma) d\sigma &+ \frac ec \int_0^T \left(\vec A(\vec x(\sigma))\cdot \frac{d\vec x(\sigma)}{d\sigma}\right)(\sigma)d\sigma\\
&- e \int_0^T \phi(\vec x(\sigma))d\sigma \tag{16}
\end{align*}

The above heuristic (from a rigorous mathematical point of view [3]) manipulation shows our claims.

\vglue .3in

\noindent{\large\bf 3.\, Conclusions}

\bigskip

As a general conclusion of our note on the nature of the Feynman path integral in the presence of on external electromagnetic potential we stress that we have substituted the whole machinery of sliced steps for defining Feynman path integrals for the somewhat classical stochastic equation eq(6) driven by (still mathematically formal) a Fresnel quantum white noise $\vec n(\sigma)$. This result may be of practical use for Monte-Carlo sampling evaluations of observables since the numerical approximated solution of the quantum Fresnel stochastic eq(6) appears a less formidable task than evaluating the Feynman path integral eq(16) directly. These claims are  result that eq(5) is a first order system of usual non-linear partial differential equations and eq(6) reduces to non-linear algebraic equations for the Fourier coefficients of the expansion of the closed quantum trajectory in Fourier series (note that $\vec n(\sigma) = \sum\limits_{n=1}^\infty \vec{e}_n\,{\rm sin}\left(\dfrac{2n\pi}{T}\,\sigma\right))\cdot$

\vglue .1in

\noindent{\large\bf Acknowledgements:} We are thankfull to CNPq for a fellowship and thankfull also to Professor José Helayel - CBPF and Professor W. Rodrigues - IMEC-UNICAMP.

\vglue .2in

\noindent{\large\bf References}
\begin{itemize}
\item[{[1]}] Luiz C.L. Botelho - Modern Physics Letters B, vol. 14, n.3, 73-78, (2000).
\item[{-}] Luiz C.L. Botelho - Inst. J. Theor. Phys., v 1, 10-15 (2016).
\item[{[2]}] L.S. Schulman - Techniques and Applications of Path Integration - John Wiley \& Sons (1981).
\item[{[3]}] Luiz C.L. Botelho - A note on Feynman-Kac path integral representations for scalar wave actions - Random Operators and Stochastic Equations, V.21, pp. 271-292, 2013. DOI: 10.1515/rose-2013-012.
\item[{[4]}] Luiz C.L. Botelho - Il Nuovo Cimento, vol. 117 B, 37-55. N.1, Gennaro 2002.
\item[{-}] Luiz C.L. Botelho - Modern Physics Letters B, vol. 12, 
N. 14815, 569-523, (1998).
\item[{-}] Luiz C.L. Botelho - Phys. Rev. 58 E, 1141-1143, Jul, 1998.
\item[{-}] Luiz C.L. Botelho - Modern Physics Letters 16 B, 793-806, Sep 2002.
\end{itemize}

\end{document}